\documentclass[aps, pra,twocolumn,showpacs,superscriptaddress,10pt]{revtex4-1}

\usepackage{amsmath}
\usepackage{amsfonts}
\usepackage{amssymb}
\usepackage{graphicx}
\usepackage{bm}

\newtheorem{propo}{Proposition}

\newcommand{\ket}[1]{\left | #1 \right\rangle}

\newcommand{\Tr}{\mathrm{Tr}}

\renewcommand{\epsilon}{\varepsilon}

\begin{document}

\title{Correlation tensor criteria for genuine multiqubit entanglement}

\author{Wies{\l}aw Laskowski}
\affiliation{Institute of Theoretical Physics and Astrophysics, University of Gda\'nsk, 80-952 Gda\'nsk, Poland}

\author{Marcin Markiewicz}
\affiliation{Institute of Theoretical Physics and Astrophysics, University of Gda\'nsk, 80-952 Gda\'nsk, Poland}

\author{Tomasz Paterek}
\affiliation{Centre for Quantum Technologies, National University of Singapore, 117543 Singapore, Singapore}

\author{Marek \.Zukowski}
\affiliation{Institute of Theoretical Physics and Astrophysics, University of Gda\'nsk, 80-952 Gda\'nsk, Poland}

\date{\today}

\begin{abstract}
We present a development of a geometric approach to entanglement indicators. The method is applied to detect genuine multiqubit entanglement. The criteria are given in form of non-linear conditions imposed on correlation tensors. Thus they involve directly observable quantities, and in some cases require only few specific measurements to find multiqubit entanglement. The non-linearity of each of the criteria
allows detection of entanglement in wide classes of states. In contrast to entanglement witnesses, which in the space of Hermitian operators define a hyperplane, the new conditions define a geometric figure encapsulating the non-fully entangled states within it.    
\end{abstract}


\maketitle

\section{Introduction}

Since quantum entanglement is both a basic resource  in  quantum communication and  quantum information processing, and a fundamental phenomenon in considerations related to foundations of quantum physics, its qualitative and quantitative characterization becomes of a great importance for practical as well as purely theoretical reasons \cite{HHHH09, PAN}.

In contrast to the bipartite case, where the structure of entanglement is very simple (the state is either entangled or separable), 
for many subsystems the characterization of entanglement becomes more complex due to many possible ways of partitioning the whole system into subsystems.
Several different approaches to detect genuine multipartite entanglement have been proposed: 
based on Bell inequalities \cite{CGPRS02, SU02, LZ05, SKLWZW08, SU08}, 
using entanglement witnesses \cite{GT09, BJ11}, 
based on relations between elements of density matrices \cite{GS10, HMGH10}, 
utilizing Fisher information \cite{ HLKSWWPS, T} 
and finally using correlation tensors \cite{VH11}.

In this contribution we further develop geometric approach to entanglement detection proposed in Ref. \cite{BBLPZ08}.
We show how this approach leads to necessary and sufficient conditions for various forms of multipartite entanglement and derive explicit criteria for different types of non-separability.
The resulting conditions are of the form of non-linear combinations of correlation functions 
and sometimes the criteria are very simple and require only limited number of specific directly measurable data.
The non-linearity of our criteria makes them often more versatile than entanglement witnesses.
We provide examples in which our criterion detects genuine multipartite entanglement of different families of quantum states.
In our investigations, an inspiring role played the work of Yu et al. \cite{YU}, in which a general non-linear condition for two qubit entanglement was found.

\section{Representation of states in terms of correlations}

Any $n$-qubit state can be expressed as
\begin{equation}
\rho=\frac{1}{2^n}\sum_{\mu_1,...,\mu_n=0,1,2,3}T_{\mu_1,..., \mu_n}\sigma_{\mu_1}\otimes...\otimes \sigma_{\mu_n},
\label{cortensor}
\end{equation}
where $\sigma_{\mu_k} \in \{\openone, \sigma_1, \sigma_2, \sigma_3 \}$ are the Pauli matrices of the $n$th observer.
The coefficients $T_{\mu_1,...,\mu_n}$ are real numbers in $[-1,1]$ given by correlation function values for measurements of products of Pauli operators
\begin{equation}
T_{\mu_1,...,\mu_n}=\langle \sigma_{\mu_1}\otimes...\otimes \sigma_{\mu_n} \rangle_{\rho}=\textrm{Tr} \left( \rho\, \sigma_{\mu_1}\otimes...\otimes \sigma_{\mu_n} \right).
\end{equation}
A specific role is played by the components which involve only indices $1,2,3$ (such indices will be denoted by Latin letters). 
The quantity: 
\begin{equation}
\hat T \equiv \sum_{i_1,\ldots,i_n=1}^3 T_{i_1,..., i_n} e^{i_1}\otimes...\otimes e^{i_n},
\end{equation}
where $\{e^{i_m}\}_{i_m=1}^3$ is a basis in $\mathbb R^3$, transforms like a tensor under local unitary transformations on the qubits.
As a consequence, we refer to it as a correlation tensor of the state $\rho$. 
The whole object $T \equiv  T_{\mu_1,..., \mu_n} $, where the indices take on values $\mu_k=0,1,2,3,$ will be called an extended correlation tensor. 
Its components with $k$ zeros are $(n-k)$-rank tensors.
Extended correlation tensors belong to a real vector space with a  natural scalar product
\begin{equation}
(X,Y)=\sum_{\vec \mu}X_{\vec \mu}Y_{\vec \mu},
\end{equation}
where  $\vec \mu=\left(\mu_1,...,\mu_n\right)$ and $\mu_k=0,1,2,3$.
We can generalize the notion of a scalar product between correlation tensors introducing a positive semidefinite metric $G$. 
A generalized scalar product has then the following form:
\begin{equation}
\left(X,Y\right)_{G}=\sum_{\vec \mu,\vec \nu} X_{\vec \mu}G_{\vec \mu \vec \nu}Y_{\vec \nu}.
\label{dotpr}
\end{equation}
This scalar product induces a $G$-norm:
\begin{equation}
||T||_{G}^2 = (T,T)_{G}.
\label{gnorm}
\end{equation}
Such a generalized scalar product and $G$-norm were used in \cite{BBLPZ08}.

\section{Multipartite entanglement}

Let us begin with a classification of entanglement of multipartite states.

A pure $n$-partite state $|\psi\rangle$ is called $k$-product, if it can be represented as a tensor product of $k$ pure $i_m$-partite states:
\begin{equation}
|\psi_{k\mathrm{-prod}}\rangle=|\psi_{i_1}\rangle\otimes\ldots\otimes|\psi_{i_k}\rangle,
\label{kprodukt}
\end{equation}
where of course, $\sum_{m=1}^k i_m=n$.
There can be different types of $k$-product states corresponding to different ways of partitioning $n$ into a sum of $k$ integers.
We will refer to a  definite type of $k$-product state as $(i_1+\ldots+i_k)$-partition product state.
Note that different $k$-product states of the same type may involve in their partitions different physical subsystems.
For example, $(2+1)$-partition product states of three particles $A$, $B$ and $C$ are: $\ket{\psi_A} \ket{\psi_{BC}}$, $\ket{\psi_B} \ket{\psi_{AC}}$ and $\ket{\psi_C} \ket{\psi_{AB}}$. We will refer to all this types of states as: $(A+BC)$, $(B+AC)$ and $(AB+C)$ - product states respectively.

The $n$-partite state $\rho$ is called $k$-separable, if it can be expressed as a probabilistic (convex) mixture of pure $k$-product states:
\begin{equation}
\rho_{k\mathrm{-sep}}=\sum_i p_i |\psi^i_{k\mathrm{-prod}}\rangle\langle\psi^i_{k\mathrm{-prod}}|.
\label{romixt}
\end{equation}
In this terminology fully separable states are $n$-separable.
If a state is not $k$-separable, then it must involve entanglement between at least $n-k+2$ parties.
Accordingly, a state is called genuinely multipartite entangled, if it is not biseparable, i.e. not $2$-separable.

\begin{figure}[!b]
\begin{center}
\includegraphics[scale=0.4]{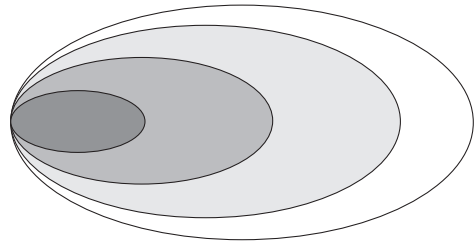}
\end{center}
\caption{Sets of $k$-separable states: all sets are convex, 
the darker is the color of a set the more separable are its states, i.e. a set of $k$-separable states contains as its subsets all more-than-$k$-separable states.
}
 \label{FIG_K_SEP}
\end{figure}

Clearly, a set of $k$-separable states is convex and it is a subset of $(k-1)$-separable states as illustrated in Fig. \ref{FIG_K_SEP}.
They also touch each other in the following meaning:  infinitesimally close to fully separable states
there are states with entanglement between arbitrary number of subsystems.  For an illustrative example see Section \ref{SEC_GHZ_METRIC}.

\section{Necessary and sufficient conditions for multipartite entanglement}

To indicate a case of a non-$k$-separability we shall use the following simple geometrical observation, which is a corollary of the results of  \cite{BBLPZ08}, namely that:
\begin{equation}
\left(\exists_G \max_{T^{k\textrm{-}sep}}(T^{k\textrm{-}sep},T)_G< ||T||_G^2\right)\Longrightarrow T \textrm{ is not }k\textrm{-sep. },
\label{main ineq}
\end{equation}
where $T^{k\textrm{-}sep}$ is a correlation tensor of some $k$-separable state, and $G$ is a metric in the sense of equation (\ref{dotpr}).
It forms a sufficient condition for multipartite entanglement between at least $n-k+2$ parties.

The advantage of using scalar products to entanglement detection is that optimization over separable states in (\ref{main ineq})
can be replaced with optimization over pure $k$-product states only:
\begin{eqnarray}
\max_{T^{k\textrm{-}sep}}(T^{k\textrm{-}sep},T)_G&=&\max_{\{\{p_i\},T^{k\textrm{-}prod}\}}\left(\sum_i p_i T^{k\textrm{-}prod}_{(i)},T\right)_G \nonumber\\
&\leq& \max_{T^{k\textrm{-}prod}}(T^{k\textrm{-}prod},T)_G,
\end{eqnarray}
This follows directly from linearity of a scalar product and convexity of $k$-separable states.
Since $k$-separable states may involve different types of $k$-product states, one should optimize over all possible partitions compatible with $k$-separability.
This reveals another feature of our condition of potential practical value:
for different partitions $\pi$ one can use different metrics to reveal that the state is not $k$-product.
All above implies the following modified condition:
\begin{eqnarray}
\left(\forall_{\pi}  \exists_{G_{\pi}}  \max_{T^{k\textrm{-}prod}_{\pi}}(T^{k\textrm{-}prod}_{\pi},T)_{G_{\pi}}< ||T||_{G_{\pi}}^2\right)&\Longrightarrow& 
 \nonumber\\
  T \textrm{ is not }k\textrm{-separable} .& &
\label{main ineq sigma}
\end{eqnarray}

Finally, it is easy to show by adapting the reasoning of \cite{BBLPZ08}, that in case of any non-$k$-separable multiqubit state, for any given partition, one can find such a metric $\tilde G$ that LHS of the inequality in condition (\ref{main ineq sigma}) holds, which leads to a necessary and sufficient condition for genuine multipartite entanglement.
Indeed, the condition  of rejecting full separability, originally formulated  in \cite{BBLPZ08} in terms of density matrices, can be put as follows:
\begin{equation}
\max_{\rho_{prod}} \operatorname{Tr}(\rho \tilde G \rho_{prod})<\operatorname{Tr}(\rho \tilde G \rho)\Longrightarrow \rho \textrm{ is not product state},
\label{main ineq prod dens}
\end{equation}
where now $\tilde G$ is a positive semidefinite superoperator. In \cite{BBLPZ08} it is shown, that, conversly, if a state $\rho$ is not separable, there exists a positive semidefinite superoperator $\tilde G$, such that the inequality (\ref{main ineq prod dens}) holds.
To state this fact clearly let us introduce new notation. Let $\{f_{\mu}\}_{\mu=0}^3$ denote standard basis in the space of $2 \times 2$ matrices  over complex numbers $\mathbb M_{2}(\mathbb C)$, that is the space of operators acting on a Hilbert space $\mathbb C^{2}$ of a single qubit. Any density matrix $\rho$ can then be decomposed as $\rho=\sum_{\vec \mu} \rho_{\vec \mu} f_{\vec \mu}$, where $\vec \mu=\{\mu_1,\ldots,\mu_n\}$ and $f_{\vec \mu}=f_{\mu_1}\otimes\ldots\otimes f_{\mu_n}$ is a basis in a tensor product space $(\mathbb M_2)^{\otimes n}=\mathbb M_{2^n}$. 
According to \cite{BBLPZ08}, if a state $\rho$ is not separable, then there exists a superoperator $\tilde G$, such that:
\begin{equation}
\label{prop proj G}
\max_{\rho_{prod}} \operatorname{Tr}(\rho \tilde G \rho_{prod})<\operatorname{Tr}(\rho \tilde G \rho).
\end{equation}
In the case of rejecting $k$-separability instead of full separability we proceed in full analogy, except for the fact that for optimilizing the LHS of (\ref{main ineq prod dens}) for each partition, we treat all elements of this partition as a single system (of course it can be of higher dimension). Having $k$ single systems we can find a metric $\tilde G$ for each partition separately, since correctness of condition (\ref{prop proj G}) does not depend on dimensions of single systems under consideration.
By choosing a specific basis (putting $f_{\mu}=\sigma_{\mu}$), all above considerations can be translated into correlation tensor representation of a state (\ref{cortensor}).  Indeed, having a density matrix of any state in correlation tensor form (\ref{cortensor}), we can find its components in a standard basis in the following manner:
\begin{equation}
\rho_{\vec \mu}=\sum_{\vec \nu} T_{\vec \nu} U_{\vec \mu \vec \nu}.
\label{st into cor}
\end{equation}
The scalar product between density matrices $\rho$ and $\rho'$ in metric $\tilde G$ can now be reformulated as a scalar product between extended correlation tensors:
\begin{eqnarray}
\operatorname{Tr}(\rho \tilde G \rho') &=& (\rho,\rho')_{\tilde G} = \sum_{\vec \mu \vec \nu} \rho_{\vec \mu} \tilde G_{\vec \mu \vec \nu} \rho'_{\vec \nu}\nonumber\\
&=&\sum_{\vec \mu \vec \nu} \sum_{\vec \gamma \vec \delta} T_{\vec \gamma}U_{\vec \mu \vec \gamma} \tilde G_{\vec \mu \vec \nu} T'_{\vec \delta} U_{\vec \nu \vec \delta}\nonumber\\
&=&\sum_{\vec \gamma \vec \delta} T_{\vec \gamma} G_{\vec \gamma \vec \delta} T'_{\vec \delta }=(T,T')_G .
\label{main transf}
\end{eqnarray}
Above transformation implies, that the metric operator $G$ which defines a scalar product of correlation tensors (cf. (\ref{dotpr})) corresponding to a scalar product of density matrices with superoperator $\tilde G$ has the following form:
\begin{equation}
G_{\vec \gamma \vec \delta}= \sum_{\vec \mu \vec \nu}  U_{\vec \mu \vec \gamma} \tilde G_{\vec \mu \vec \nu}U_{\vec \nu \vec \delta}
\label{GGtilde}
\end{equation}

Altogether we have shown the following proposition fully characterizing $k$-separability:

\begin{propo}
An $n$-particle state endowed with extended correlation tensor $T$ is not $k$-separable 
if and only if for every partition $\pi$ into $k$ subsystems there exists a metric $G_{\pi}$,
such that the following inequality holds:
\begin{equation}
\max_{ T^{k\textrm{-prod}}_\pi } (T^{k\textrm{-prod}}_\pi,T)_{G_{\pi}} <  (T,T)_{G_{\pi}}.
\end{equation}
\label{IFF}
\end{propo}

\section{Examples}

We present here a few examples of applications of Proposition \ref{IFF}.
It leads to simple sufficient conditions for multipartite entanglement that are sometimes also necessary and
detects entanglement of various classes of states (which is impossible using entanglement witnesses).

\subsection{Three-qubit entanglement}

To reveal a genuine 3-partite entanglement in a three qubit state we have to exclude the case of biseparability. 
We derive several sufficient criteria using different metric tensors.

\subsubsection{Standard metric}
Here we shall give a condition which is unbiased in its formulation with respect to any family of entangled states.
It uses the diagonal metric, that with $G_{\vec{\mu}\vec{\nu}}$ in the form of a  Kronecker delta $\delta_{\vec{\mu}\vec{\nu}}$.

Let $\hat T$ be a correlation tensor of a $3$-qubit state, and let $\hat T^{2+1}$ be a correlation tensor of a $(2+1)$-partition product $3$-qubit state. Assuming standard (Euclidean) scalar product in space  $\mathbb R^9$:
\begin{equation}
(\hat X,\hat Y)=\sum_{i,j,k=1}^3 X_{ijk} Y_{ijk},
\end{equation}
 we show, that $\max_{\sigma}\max_{\hat T^{2+1}_{\sigma}}(\hat T^{2+1}_{\sigma},\hat T),$
 where $\sigma$ denotes permutation of biproduct states (here three possible splittings),
  is upper bounded by:
\begin{equation}
\max_{\sigma}\max_{\sigma(\hat O \otimes \hat O', \hat \openone)}\sqrt{\sum_{i=1}^3\left(\left|T_{\sigma(11i)}-T_{\sigma(22i)}\right|+|T_{\sigma(33i)}|\right)^2},
\label{eq2p}
\end{equation}
where $\sigma(\hat O \otimes \hat O', \hat \openone)$ means, that the second maximization is done over local orthogonal transformations applied to subsystems over which, for a given $\sigma$, summation is not performed, e.g. if $\sigma$ is a trivial permutation, $\sigma(\hat O \otimes \hat O', \hat \openone)=\hat O \otimes \hat O' \otimes \hat \openone$.

Clearly $\hat T^{2+1}=\hat T^2\otimes \hat T^1$. Any pure two qubit state can be expressed in a Schmidt basis in the form: $\cos \theta |00\rangle+\sin \theta |11\rangle$, which has the following  nonvanishing terms of correlation tensor $\hat T^2$:
\begin{eqnarray}
T_{11}&=&\sin 2\theta \nonumber\\
T_{22}&=&-\sin 2\theta \nonumber\\
T_{33}&=&1
\label{eq3a}
\end{eqnarray}
while $\hat T^1$ is a Bloch vector:
\begin{equation}
\hat T^1=\vec m=[m_1, m_2, m_3], \quad \textrm{with} \quad \sqrt{m_1^2+m_2^2+m_3^2}=1.
\end{equation}
Hence the scalar product $(\hat T^{2+1}_{\sigma},\hat T)$, where $\sigma$ denotes some permutation of indices defining to which subsystems tensors  $\hat T^{2}$ and $\hat T^{1}$ correspond, can be expressed as:
\begin{equation}
(\hat T^{2+1}_{\sigma},\hat T)=\sum_{i=1}^3 \left((T_{\sigma(11i)}-T_{\sigma(22i)})\sin 2\theta +T_{\sigma(33i)}\right)m_i.
\label{eq4}
\end{equation}
Since $\vec m$ is arbitrary unit vector, the maximization over $\vec m$ gives:
\begin{eqnarray}
& &\max_{\vec m}(\hat T^{2+1}_{\sigma},\hat T)=\nonumber\\
&=&\sqrt{\sum_{i=1}^3\left(\left(T_{\sigma(11i)}-T_{\sigma(22i)}\right)\sin 2\theta +T_{\sigma(33i)}\right)^2}.\nonumber\\
\label{eq5}
\end{eqnarray}
Finally we have to maximize over $\theta$, over permutations $\sigma$ of subsystems  and over all possible local rotations $\hat O \otimes\hat O'$  applied to subsystems, over which we do not sum in Eq. (\ref{eq5}). At this stage we do not need to maximize over rotations applied to the subsystem over which we sum in the above equation, since maximization over a unit $\vec m$ is equivalent to maximization over all possible rotations of this vector, and the maximization over rotations of the scalar product $(\hat T^{2+1}_{\sigma},\hat T)$ can be performed over any of the tensors in this product.

Adopting the notation introduced in Eq. (\ref{eq2p}) 
and using the fact that for reals $r_1$ and $r_2$ we have $(r_1 \sin 2 \theta + r_2)^2 \le (|r_1| + |r_2|)^2$, 
we reach the following condition:
\begin{propo}
If the following inequality holds:
\begin{equation}
\max_{\sigma,\sigma(\hat O \otimes \hat O', \hat \openone)}\sqrt{\sum_{i=1}^3\left(\left|T_{\sigma(11i)}-T_{\sigma(22i)}\right|+|T_{\sigma(33i)}|\right)^2}< ||\hat T||^2, 
\label{ineqp}
\end{equation}
then the state described by correlation tensor $\hat T$ is genuinely 3-partite entangled.
\label{prop1}
\end{propo}
However, as  the left-hand side of this proposition may be strictly bigger than Eq. (\ref{eq5}), in some cases it is more effective to directly use condition
\begin{equation}
\max_{\sigma} \max_{\hat T^{2+1}_{\sigma}} (\hat T^{2+1}_{\sigma}, \hat T) \leq ||\hat T||^2.
\end{equation}
For example, for a generalized $3$-partite GHZ state:
\begin{equation}
|GHZ_{\alpha}\rangle=\cos \alpha |000\rangle+\sin \alpha |111\rangle,
\label{ghza}
\end{equation}
mixed with white noise:
\begin{equation}
v |GHZ_{\alpha}\rangle\langle GHZ_{\alpha}|+(1-v)\frac{1}{8}\openone,
\label{eq6}
\end{equation}
Eq. (\ref{eq5}) is maximized for vanishing $\theta$, and reads $v\sqrt{1+3 \sin^2 2\alpha}$.
The same value is obtained after applying local rotations as we verified numerically.
Since the squared length of the correlation tensor equals $v^2 (1+3 \sin^2 2\alpha)$,
we find that if $v$ exceeds the critical value
\begin{equation}
v_{crit}=\frac{1}{\sqrt{1+3 \sin^2 2\alpha}},
\label{V_CRIT_GGHZ}
\end{equation}
the state is genuinely $3$-partite entangled for any $\alpha$.
Local rotations applied to the left-hand side of (\ref{ineqp}) make its value higher than the one which follows from Eq. (\ref{eq5}),
and accordingly Proposition \ref{prop1} does not detect as many states as direct application of Eq. (\ref{eq5}).
Finally, note that the critical visibility (\ref{V_CRIT_GGHZ}) holds for all possible locally unitarily equivalent $3$-partite GHZ states.
For a symmetric GHZ state (that is for $\alpha=\pi/4$), $v_{crit}=\frac{1}{2}$.
Note that even for an arbitrarily small but finite $\alpha$, the state (\ref{ghza}) is $3$-partite entangled, while for $\alpha=0$ it is fully separable. 

In case of a 3-partite W state $|W_3\rangle=\frac{1}{\sqrt{3}}(|100\rangle+|010\rangle+|001\rangle)$ mixed with white noise:
\begin{equation}
\rho_W (p)=v|W_3\rangle\langle W_3|+(1-v) \frac{1}{8} \openone
\label{noisy W}
\end{equation}
the critical visibility for detection of genuine 3-qubit entanglement with the condition (\ref{ineqp}) has been calculated numerically, and is equal to $v_{crit}\approx0.636$.

We stress that this approach is quite versatile, it allows to detect three-particle entanglement despite the fact that GHZ and W states are of a different nature.

\subsubsection{Modified metric and generalized Schmidt decomposition of the correlation tensor}

The criterion in Proposition \ref{prop1} can be made more efficient and simplified by changing a metric and applying a generalized Schmidt decomposition \cite{SUD00} to the correlation tensor. 

In order to arrive at a simple criterion, we modify the metric (hence also scalar product in the sense of Eq. (\ref{dotpr})) to get rid of terms of the $T_{\sigma(33i)}$ type.
Thus, we put
\begin{equation}
||\hat T||^2_{mod}=\sum_{i,j,k=1}^3 T_{ijk}^2-\sum_{l=1}^3 T_{\sigma(33l)}^2.
\label{eq7}
\end{equation}
Using this metric, the inequality (\ref{ineqp}) can be rewritten as:
\begin{equation}
\max_{\sigma}\max_{\sigma(\hat O \otimes \hat O', \hat \openone)}\sqrt{\sum_{i=1}^3\left(T_{\sigma(11i)}-T_{\sigma(22i)}\right)^2}< ||\hat T||^2_{mod}.
\label{ineq1}
\end{equation}
Since the condition (\ref{main ineq}) is valid in any metric $G$, we have the following modified criterion:
\begin{propo}
If the following inequality holds:
\begin{equation}
\max_{\sigma}\max_{\sigma(\hat O \otimes \hat O', \hat \openone)}\sqrt{\sum_{i=1}^3\left(T_{\sigma(11i)}-T_{\sigma(22i)}\right)^2}< ||\hat T||^2_{mod}, 
\label{ineq2}
\end{equation}
where $||\hat T||_{mod}^2=\sum_{i,j,k=1}^3 T_{ijk}^2-\sum_{l=1}^3 T_{\sigma(33l)}^2$, 
then the state described by correlation tensor $\hat T$ is genuinely 3-partite entangled.
\label{prop2}
\end{propo}

We can further simplify this condition applying handy features of a generalized Schmidt decomposition to the correlation tensor \cite{SUD00}. According to Theorem 1 in  \cite{SUD00}, for any tensor 
\begin{equation}
\hat T=T_{i_1,..., i_n} e^{1}_{i_1} \otimes \dots \otimes e^{n}_{i_n}, \quad \textrm{with} \quad i_k=1,\ldots,d,
\end{equation}
where $\{e^{m}_{i_m}\}_{i_m=1}^d$ is a basis in some $d$-dimensional vector space,
there exists a basis $s^{1}_{i_1}\otimes...\otimes s^{n}_{i_n}, i_k=1,\ldots,d$, which we shall call a generalized Schmidt basis, in which the components $T'_{i_1,..., i_n} $ of  tensor $\hat T$:
\begin{equation}
\hat T=T'_{i_1,..., i_n} s^{1}_{i_1}\otimes...\otimes s^{n}_{i_n}
\end{equation}
have the following properties:
\begin{eqnarray}
\label{genschmidt1}
&\bullet& T'_{\sigma(j,i,\ldots,i)}=0 \textrm{ for } 1\leq i < j \leq d,\\
\label{genschmidt2}
&\bullet& T'_{i_1,..., i_n} \textrm{is non-negative if}\nonumber\\
&&\textrm{at most one of the } i_k \textrm{ differs from } d,\\
\label{genschmidt3}
&\bullet& |T'_{j,\ldots,j}| \geq |T'_{i_1,..., i_n}| \textrm{ if } j\leq i_r \textrm{ for all } r = 1, ..., n.
\end{eqnarray}
Assume that the correlation tensor $\hat T$ in (\ref{ineq2}) is expressed in a generalized Schmidt basis. Then the property (\ref{genschmidt1}) implies, that 
 from among the following two groups of terms of the correlation tensor:
$$\{T_{111},T_{112}, T_{113}\}$$
$$\{T_{221},T_{222}, T_{223}\}$$
only one term in each group is nonzero. Let us assume without loosing generality that $T_{111}$ is the maximal generalized Schmidt coefficient and only $T_{221}$ is non-zero. This implies that for all $\sigma$:
$$\sqrt{\sum_{i=1}^3\left(T_{\sigma(11i)}-T_{\sigma(22i)}\right)^2}\leq  |T_{111}| +|T_{221}| \leq 2|T_{111}|.$$
This property gives the following criterion:
\begin{propo}
If the following inequality holds:
\begin{equation}
||\hat T||^2_{mod}> 2T_{max},
\label{ineq3}
\end{equation}
where $||\hat T||_{mod}^2=\sum_{i,j,k=1}^3 T_{ijk}^2-\sum_{l=1}^3 T_{\sigma(33l)}^2$, and $T_{max}$ is the maximal possible value a correlation tensor element for the given state, 
then the state described by correlation tensor $\hat T$ is genuinely $3$-partite entangled.
\label{prop3}
\end{propo}
The above condition can be simplified further to a weaker one:
$$||\hat T||^2_{mod}>2,$$
which detects a smaller class of entangled states, but is experimentally very handy. One can measure components of $\hat T$ which enter $||\hat T||^2_{mod}$, and once the sum  (\ref{eq7}) is above $2$, a genuine three particle entanglement is confirmed.

All above analysis can be performed in full analogy in case of $4$-partite states. Due to complexity of formulas, conditions for $4$-qubit entanglement are presented in Appendix A.

\subsection{GHZ metric}
\label{SEC_GHZ_METRIC}
In this section we shall study an approach that favours a certain family of states, namely the GHZ ones. 

Consider the so-called Greenberger-Horne-Zeilinger state of $n$ qubits:
\begin{equation}
\ket{GHZ} = \frac{1}{\sqrt{2}} \left( \ket{0 \dots 0} + \ket{1 \dots 1} \right).
\label{GHZ_STATE}
\end{equation}
It has the following non-vanishing elements of extended correlation tensor \cite{LPZB2010}:
\begin{eqnarray}
T_{\underbrace{y...y}_{2k}x....x}^{\mathrm{GHZ}} & = & (-1)^{k}, \quad k = 0,1,...,\lfloor \tfrac{n-1}{2} \rfloor \nonumber \\
T_{\underbrace{z...z}_{2k}0...0}^{\mathrm{GHZ}} & = & 1,
\end{eqnarray}
and with indices permuted.
In turns out that Proposition \ref{IFF} with a diagonal metric $G$,
\begin{equation}
(X,Y)_G = \sum_{\vec \mu} X_{\vec \mu} G_{\vec \mu} Y_{\vec \mu},
\end{equation}
for the specific case of  $G_{\vec \mu} = |T_{\vec \mu}^{\mathrm{GHZ}}|$  and additionally putting $G_{0,\dots,0} = 0$,
leads to optimal detection of a genuine multipartite entanglement of noisy GHZ states
\begin{equation}
\rho = v |GHZ\rangle\langle GHZ|+(1-v)\frac{1}{2^n}\openone.
\end{equation}

Using the metric, the right-hand side of Proposition \ref{IFF} for state $\rho$ is given by $(2^{n}-1) v^2$.
In order to find the maximum of the left-hand side over bi-product states, we write
\begin{equation}
L \equiv (T,T^{\mathrm{bi-prod}})_G = v \sum_{\vec \mu \in \mathcal{GHZ}} \textrm{sgn}(T^{\mathrm{GHZ}}_{\vec \mu}) T^{\mathrm{bi-prod}}_{\vec \mu},
\end{equation}
where we introduce a convenient notation for summing over non-zero elements of a ''GHZ'' metric, 
and note that after transforming correlation tensors to density operators this reads
\begin{eqnarray}
L/v & = & 2^{n-1} (\rho_{1,1}+\rho_{1,2^n}+\rho_{2^n,1}+\rho_{2^n,2^n}) -1 \\
& = & 2^{n} \Tr(\rho^{\mathrm{GHZ}} \rho^{\mathrm{bi-prod}})-1.
\end{eqnarray}
In the last step we use $\Tr(\rho^{\mathrm{GHZ}} \rho^{\mathrm{bi-prod}}) \le \frac{1}{2}$, which holds for all bi-separable states \cite{ABLS01}.
Therefore, $L \le (2^{n-1}-1) v$ and the state $\rho$ is shown to be genuinely multipartite entangled if
\begin{equation}
v > \frac{2^{n-1}-1}{2^n-1},
\label{V_GHZ}
\end{equation}
which is known to be optimal \cite{GS10,GHH10}.

The same metric can reveal genuine multipartite entanglement of many other states.
Consider generalized GHZ states
\begin{equation}
\ket{GHZ_\alpha} =  \cos \alpha \ket{0 \dots 0} + \sin \alpha \ket{1 \dots 1},
\end{equation}
with $\alpha \in [0,\frac{\pi}{4}]$.
The non-vanishing components of its correlation tensor are given by permutations of indices of the following ones
\begin{eqnarray}
T_{\underbrace{y...y}_{2k}x....x} & = & (-1)^{k} \sin 2 \alpha, \quad k = 0,1,...,\lfloor \tfrac{N-1}{2} \rfloor \nonumber \\
T_{\underbrace{z...z}_{k}0...0} & = & 
\Bigg\{ \begin{array}{ll}
1 & \textrm{ for } k \textrm{ even}, \\
\cos 2 \alpha & \textrm{ for } k \textrm{ odd}. \\
\end{array}
\end{eqnarray}
Taking again the ''GHZ'' metric, one can repeat the proof which led to (\ref{V_GHZ})
with the only difference that now $$\Tr(\rho^{\mathrm{GHZ_\alpha}} \rho^{\mathrm{bi-prod}}) \le \cos^2 \alpha .$$
In this way we obtain that generalized GHZ state mixed with white noise is genuinely multipartite entangled for
\begin{equation}
v > \frac{2^{n}\cos^2 \alpha - 1}{2^{n} - 1}.
\end{equation}
Finally, note that this state is fully separable only for $\alpha = 0$. 
Already for infinitesimally small $\alpha$ it can involve entanglement between all $n$ parties.
Clearly, a similar statement would hold for a generalized GHZ state between $n-1$ parties and the $n$th party having an uncorrelated state.
Therefore, in infinitesimal neighborhood of the state $\ket{0 \dots 0}$ there are states with entanglement between an arbitrary number of subsystems.

\section{Conclusions}

We have presented several sufficient criteria for a multipartite entanglement in the form of nonlinear conditions imposed on correlations of the tested state.
The conditions are given in a convenient and simple form, and can be directly applied to given families of entangled states. 

An important advantage of our criteria is that in many cases only few definite measurements suffice to detect multiqubit entanglement. 

Presented criteria are more general than entanglement witnesses due to their nonlinearity.
A single new criterion detects a genuine entanglement of many different families of states, whereas one definite witness can detect entanglement of one family of states only.

Here we have given only several examples, however one can construct infinitely many other ones. Note, that only our conditions  with the GHZ metric involved correlations of all qubits as well as only some of them. Note that this is the case for the universal two qubit entanglement condition given in \cite{YU}, thus this seems to be a promising direction of a further research. 
Different series of conditions of such a kind, with surprising properties,  will be presented elsewhere \cite{US}.

\section{Acknowledgements}

We thank Piotr Badzi{\c{a}}g and Marcus Huber for stimulating discussions.

This work is supported by the EU program Q-ESSENCE (Contract
No.248095), the MNiSW Grant no. N202 208538, 
and the National Research Foundation and Ministry of Education in Singapore.

The contribution of MM is supported within the International PhD Project
``Physics of future quantum-based information technologies''
grant MPD/2009-3/4 from Foundation for Polish Science and  by the University of Gda\'nsk grant 538-5400-0623-1.

WL acknowledges financial support from European Social Fund as a
part of the project ``Educators for the elite - integrated training
program for PhD students, post-docs and professors as academic
teachers at University of Gdansk'' within the framework of Human
Capital Operational Programme, Action 4.1.1, Improving the quality of
educational offer of tertiary education institutions.

\appendix
\section{}

\subsection{Genuine $4$-partite entanglement in four-qubit states}

\subsubsection{Exclusion of  biseparability}

To exclude a biseparability of a $4$-qubit state, we have to verify the condition (\ref{main ineq}) for the case of maximizing over $(3+1)$- and $(2+2)$-partition product states:
\begin{eqnarray}
\Big( \max_{T^{3+1}}(T^{3+1},T)_G &<& ||T||^2_G \Big) \textrm{ and } \Big(\max_{T^{2+2}}(T^{2+2},T)_G < ||T||^2_G \Big) \nonumber\\
&&\Longrightarrow T \neq T^{bisep},
\label{e4q1}
\end{eqnarray}
where $G$ denotes a metric operator in a vector space in which $4$-qubit correlation tensors are embedded.
Using a version of a ''GHZ'' metric  for a four-qubit system, in which only $T_{\sigma(1122)}$-type terms occur:
\begin{eqnarray}
||\hat T||^2_{GHZ} & \equiv & T_{1111}^2+T_{1122}^2+T_{1221}^2+T_{2211}^2\nonumber\\
&+&T_{1212}^2+T_{2121}^2+T_{2112}^2+T_{2222}^2,
\label{ghzmetricxy}
\end{eqnarray}
we calculate the first term in (\ref{e4q1}).
Since $\hat T^{3+1}$ is a pure state, we have $\hat T^{3+1}=\hat T^3 \otimes \hat T^1$.
Taking $\hat T^1=\vec m=[m_1, m_2, m_3]$, with $\sqrt{m_1^2+m_2^2+m_3^2}=1$, and assuming $T^3\otimes T^1$ is (ABC + D)-type product, one obtains:
\begin{eqnarray}
(\hat  T^3\otimes \hat T^1,\hat T)&=&T_{111}m_1T_{1111}+T_{112}m_2T_{1122}\nonumber\\
&+&T_{122}m_1T_{1221}+T_{221}m_1T_{2211}\nonumber\\
&+&T_{121}m_2T_{1212}+T_{212}m_1T_{2121}\nonumber\\
&+&T_{211}m_2T_{2112}+T_{222}m_2T_{2222}.
\label{e4q2}
\end{eqnarray}
Due to properties (\ref{genschmidt1}), (\ref{genschmidt2}) and (\ref{genschmidt3}) of a generalized Schmidt decomposition \cite{SUD00} applied now to quantum states, any 3-qubit pure state can be expressed as:
\begin{eqnarray}
|\psi\rangle&=&\cos(\omega_1)|000\rangle+\cos(\omega_2)\sin(\omega_1)|001\rangle\nonumber\\
&+&e^{i\phi}\cos(\omega_3)\sin(\omega_1)\sin(\omega_2)|010\rangle\nonumber\\
&+&\cos(\omega_4)\sin(\omega_1)\sin(\omega_2)\sin(\omega_3)|100\rangle\nonumber\\
&+&\sin(\omega_1)\sin(\omega_2)\sin(\omega_3)\sin(\omega_4)|111\rangle.
\label{psi3}
\end{eqnarray}
In this parametrization, terms of $\hat  T^3$ occuring in (\ref{e4q2}) have the following form:
\begin{eqnarray}
T_{112}&=&T_{121}=T_{211}=T_{222}=0\nonumber\\
T_{111}&=&-T_{122}=-T_{221}=-T_{212}\nonumber\\
&=&\sin(2 \omega_1)\sin(\omega_2)\sin(\omega_3)\sin(\omega_4) .
\label{e4q3}
\end{eqnarray}
Hence (\ref{e4q2}) simplifies to:
\begin{eqnarray}
(\hat  T^3\otimes \hat T^1,\hat T)&=&\left(T_{1111}-T_{1221}-T_{2211}-T_{2121}\right)\nonumber\\
&&\times m_1\sin(2 \omega_1)\sin(\omega_2)\sin(\omega_3)\sin(\omega_4).\nonumber\\
\label{e4q4}
\end{eqnarray}
The maximization of (\ref{e4q4}) over $m_1, \omega_1,  \omega_2,  \omega_3,  \omega_4$ is trivial. Let us denote local orthogonal transformations $\hat O_1\otimes\hat O_2\otimes\hat O_3 \otimes\hat O_4$ as $\hat O_{tot}$. We finally obtain:
\begin{equation}
\max_{T^{3+1}}(\hat  T^{3+1},\hat T)=\max_{\hat O_{tot}}\left|T_{1111}-T_{1221}-T_{2211}-T_{2121}\right|.
\label{e4q5}
\end{equation}
In complete analogy we can find inequalities for other types of $(3+1)$-partition product states  (ABD+C, ACD+B, A+BCD), which leads to the following inequalities:
\begin{eqnarray}
\max_{\hat O_{tot}}\left|T_{1111}-T_{1221}-T_{2211}-T_{2121}\right|&<& ||\hat T||^2_{GHZ} \nonumber\\
\max_{\hat O_{tot}}\left|T_{1111}-T_{2211}-T_{1212}-T_{2112}\right|&<& ||\hat T||^2_{GHZ} \nonumber\\
\max_{\hat O_{tot}}\left|T_{1111}-T_{1122}-T_{2121}-T_{2112}\right|&<& ||\hat T||^2_{GHZ} \nonumber\\
\max_{\hat O_{tot}}\left|T_{1111}-T_{1122}-T_{1221}-T_{1212}\right|&<& ||\hat T||^2_{GHZ} .\nonumber\\
\label{ineq4q}
\end{eqnarray}
In case of pure states these inequalities allow us to check if given state is (3+1)-partition product or not:
\begin{propo}
If all the inequalities (\ref{ineq4q}) hold, then the pure state described by correlation tensor $\hat T$ is not $(3+1)$-partition product.
\label{prop4q}
\end{propo}
Now we have to calculate the second element of the conjunction in (\ref{e4q1}) involving maximization over (2+2)-partition product states.
Since 
\begin{equation}
\max_{T^{2+2}}(\hat T^{2+2},\hat T)=\max_{T^2, T^{'2}}(\hat T^{2}\otimes  \hat T^{'2},\hat T),
\label{e4q7}
\end{equation}
we need to explicitely express $\hat T^{2} \otimes\hat T^{'2}$. This is very simple due to equations (\ref{eq3a}):
\begin{eqnarray}
(\hat T^{2})_{11}&=&-\sin 2\theta \nonumber\\
(\hat T^{2})_{22}&=&\sin 2\theta \nonumber\\
(\hat T^{2})_{33}&=&1\nonumber\\
(\hat T^{'2})_{11}&=&-\sin(2\theta')\nonumber\\
(\hat T^{'2})_{22}&=&\sin(2\theta')\nonumber\\
(\hat T^{'2})_{33}&=&1.
\label{t2t2}
\end{eqnarray}
The only nonvanishing terms of the tensor $\hat T^{2} \otimes\hat T^{'2}$ are:
\begin{eqnarray}
(\hat T^{2} \otimes\hat T^{'2})_{1111}&=&\sin 2\theta \sin 2\theta' \nonumber\\
(\hat T^{2} \otimes\hat T^{'2})_{1122}&=&-\sin 2\theta \sin 2\theta' \nonumber\\
(\hat T^{2} \otimes\hat T^{'2})_{1133}&=&-\sin 2\theta \nonumber\\
(\hat T^{2} \otimes\hat T^{'2})_{2211}&=&-\sin 2\theta \sin 2\theta' \nonumber\\
(\hat T^{2} \otimes\hat T^{'2})_{2222}&=&\sin 2\theta \sin 2\theta' \nonumber\\
(\hat T^{2} \otimes\hat T^{'2})_{2233}&=&\sin 2\theta \nonumber\\
(\hat T^{2} \otimes\hat T^{'2})_{3311}&=&-\sin 2\theta' \nonumber\\
(\hat T^{2} \otimes\hat T^{'2})_{3322}&=&\sin 2\theta' \nonumber\\
(\hat T^{2} \otimes\hat T^{'2})_{3333}&=&1.
\label{t2t2a}
\end{eqnarray}
Substituting these terms one obtains (in case of maximizing over (AB+CD)-product states):
\begin{eqnarray}
&&\max_{T^2, T^{'2}}(\hat T^{2}\otimes  \hat T^{'2},\hat T)=\nonumber\\
&&\max_{\hat O_{tot}, \theta, \theta'}(\sin 2\theta \sin 2\theta' \left(T_{1111}-T_{1122}-T_{2211}+T_{2222}\right)\nonumber\\
&&+\sin 2\theta \left(T_{2233}-T_{1133}\right)\nonumber\\
&&+\sin 2\theta' \left(T_{3322}-T_{3311}\right)+T_{3333}).\nonumber\\
\label{e4q8}
\end{eqnarray}

Since only $4$ terms of (\ref{t2t2a}) occur in GHZ metric, the expression (\ref{e4q8}) has, for maximizing over (AB+CD)-type product states, the following simplified form:
\begin{eqnarray}
&&\max_{T^2, T^{'2}}(\hat T^{2}\otimes  \hat T^{'2},\hat T)=\nonumber\\
&=&\max_{\hat O_{tot},\theta, \theta'}(\sin2\theta\sin(2\theta')\left(T_{1111}-T_{1122}-T_{2211}+T_{2222}\right))\nonumber\\
&=&\max_{\hat O_{tot}}|T_{1111}-T_{1122}-T_{2211}+T_{2222}|.
\label{e4q13}
\end{eqnarray}
Taking into account other types of (2+2)-partition product states (that is of type (AC+BD) and (AD+BC)) we obtain the following set of inequalities:
\begin{eqnarray}
\max_{\hat O_{tot}}|T_{1111}-T_{1122}-T_{2211}+T_{2222}|&<& ||\hat T||^2_{GHZ}\nonumber\\
\max_{\hat O_{tot}}|T_{1111}-T_{1212}-T_{2121}+T_{2222}|&<& ||\hat T||^2_{GHZ}\nonumber\\
\max_{\hat O_{tot}}|T_{1111}-T_{1221}-T_{2112}+T_{2222}|&<& ||\hat T||^2_{GHZ}.\nonumber\\
\label{ineq5qm}
\end{eqnarray}
Finally, we obtain the Proposition, which is a direct consequence of condition (\ref{e4q1}):
\begin{propo}
If all the inequalities (\ref{ineq4q}) and (\ref{ineq5qm}) hold, then the state described by correlation tensor $\hat T$ is genuinely $4$-partite entangled.
\label{prop5q}
\end{propo}

\subsubsection{Exclusion of  $3$-separability}
Since there is only one type of $4$-partite $3$-product state, that is $(2+1+1)$-partition product, the condition (\ref{main ineq}) has the following form:
\begin{equation}
\left(\max_{T^{2+1+1}}(\hat T^{2+1+1},\hat T)_G< ||\hat T||_G^2\right)\Longrightarrow \hat T \neq \hat T^{3sep}.
\label{ineq prod 211}
\end{equation}
We proceed analogously to the case of excluding biseparability of $3$-partite state:
$\hat T^{2+1+1}=\hat T^2\otimes \hat T^1 \otimes \hat T'^1$, and we choose Schmidt basis for $\hat T^2$, in which the only nonvanishing terms are:
\begin{eqnarray}
T_{11}&=&\sin2\theta\nonumber\\
T_{22}&=&-\sin2\theta\nonumber\\
T_{33}&=&1,
\label{eq3}
\end{eqnarray}
while $\hat T^1=\vec m=[m_1, m_2, m_3]$, with $\sqrt{m_1^2+m_2^2+m_3^2}=1$, and $\hat T'^1=\vec n=[n_1, n_2, n_3]$, with $\sqrt{n_1^2+n_2^2+n_3^2}=1$.
Hence the scalar product $(\hat T^{2+1+1}_{\sigma},\hat T)$, where $\sigma$ denotes proper permutation of indices refering to subsystems, can be expressed as:
\begin{eqnarray}
&&(\hat T^{2+1+1}_{\sigma},\hat T)=\nonumber\\
&=&\sum_{i,j=1}^3 \left((T_{\sigma(11ij)}-T_{\sigma(22ij)})\sin2\theta+T_{\sigma(33ij)}\right)m_i n_i .\nonumber\\
\label{eq41}
\end{eqnarray}
Now we use Cauchy-Schwartz inequality:
\begin{eqnarray}
&&\sum_{i,j=1}^3 \left((T_{\sigma(11ij)}-T_{\sigma(22ij)})\sin2\theta+T_{\sigma(33ij)}\right)m_i n_i \leq\nonumber\\ 
&&\left(\sum_{i,j=1}^3 \left((T_{\sigma(11ij)}-T_{\sigma(22ij)})\sin2\theta+T_{\sigma(33ij)}\right)^2\right)^{\frac{1}{2}}\times\nonumber\\
&&\times\left(\sum_{i,j=1}^3 m_i^2 n_j^2\right)^{\frac{1}{2}}\leq\nonumber
\end{eqnarray}
\begin{eqnarray}
&&\left(\sum_{i,j=1}^3 \left((T_{\sigma(11ij)}-T_{\sigma(22ij)})\sin2\theta+T_{\sigma(33ij)}\right)^2\right)^{\frac{1}{2}}\times\nonumber\\
&&\times\left(\left(\sum_{i=1}^3 m_i^4\right)^{\frac{1}{2}}\left(\sum_{j=1}^3 n_j^4\right)^{\frac{1}{2}}\right)^{\frac{1}{2}}\leq\nonumber\\
&&\left(\sum_{i,j=1}^3 \left((T_{\sigma(11ij)}-T_{\sigma(22ij)})\sin2\theta+T_{\sigma(33ij)}\right)^2\right)^{\frac{1}{2}}.
\label{eq5cs}
\end{eqnarray}
The last inequality follows from Cauchy-Schwartz inequality and the fact that:
$$\sum_{i=1}^3 m_i^2\leq1 \Longrightarrow \sum_{i=1}^3 m_i^4\leq1.$$
From now on we can proceed directly as in the case of maximizing over $(2+1)$-partition product states obtaining:
\begin{propo}
If the following inequality holds:
\begin{equation}
\max_{\sigma} \max_{\hat O_{tot}}\sqrt{\sum_{i,j=1}^3\left(\left|T_{\sigma(11ij)}-T_{\sigma(22ij)}\right|+|T_{\sigma(33ij)}|\right)^2}< ||\hat T||^2 ,
\label{ineq}
\end{equation}
then the state described by correlation tensor $\hat T$ is biseparable or genuinely multiqubit entangled.
\label{prop211}
\end{propo}


\begin{thebibliography}{99}

\bibitem{HHHH09} 
R. Horodecki, P. Horodecki, M. Horodecki and K. Horodecki, 
Rev. Mod. Phys. {\bf 81}, 865–942 (2009) 

\bibitem{PAN}
J.-W. Pan, Z.-B. Chen, J.-Y. Lu, H.Weinfurter, A. Zeilinger and M. \.Zukowski, 
Rev. Mod. Phys. (in print); also e-print  arXiv:0805.2853.

\bibitem{CGPRS02}
D. Collins, N. Gisin, S. Popescu, D. Roberts, and V. Scarani, 
Phys. Rev. Lett. {\bf 88}, 170405 (2002).

\bibitem{SU02}
M. Seevinck and J. Uffink, 
Phys. Rev. A {\bf 65}, 012107 (2001).

\bibitem{LZ05}
W. Laskowski, M. \.Zukowski, 
Phys. Rev. A {\bf 72}, 062112 (2005).

\bibitem{SKLWZW08}
C. Schmid, N. Kiesel, W. Laskowski, W. Wieczorek, and M. \.Zukowski, and H. Weinfurter, 
Phys. Rev. Lett. {\bf 100}, 200407 (2008).

\bibitem{SU08}
M. Seevinck and J. Uffink, 
Phys. Rev. A {\bf 78}, 032101 (2008).

\bibitem{GT09}
O. G\"uhne and G. T\'oth, 
Phys. Rep. {\bf 474}, 1 (2009).

\bibitem{BJ11}
B. Jungnitsch, T. Moroder, O. G\"uhne, Phys. Rev. Lett. {\bf 106}, 190502 (2011).

\bibitem{HMGH10} 
M. Huber, F. Mintert, A. Gabriel, and B.C. Hiesmayr, 
Phys. Rev. Lett. {\bf 104}, 210501 (2010).

\bibitem{GS10}
O. G\"uhne and M. Seevinck, 
New J. Phys. {\bf 12}, 053002 (2010).

\bibitem{HLKSWWPS}
P. Hyllus, W. Laskowski, R. Krischek, C. Schwemmer, W. Wieczorek, H. Weinfurter, L. Pezz\'e, A. Smerzi,
arXiv:1006.4366

\bibitem{T}
Geza Toth, arXiv:1006.4368

\bibitem{VH11}
J. I. de Vicente and M. Huber, arXiv:1106.5756

\bibitem{BBLPZ08}
P. Badzi{\c a}g, {\v C}. Brukner, W. Laskowski, T. Paterek and M. \.Zukowski, 
Phys. Rev. Lett. {\bf 100}, 140403 (2008).

\bibitem{YU}
S. Yu, J.-W. Pan, Z.-B. Chen, and Y.-D. Zhang,
Phys. Rev. Lett. {\bf 91}, 217903  (2003). 

\bibitem{SUD00}
 H. A. Carteret, A. Higuchi, and A. Sudbery, 
 J. Math. Phys. {\bf 41}, 7932 (2000).

\bibitem{LPZB2010}
W. Laskowski, T. Paterek, {\v C}. Brukner, and M. \.Zukowski,
Phys. Rev. A {\bf 81}, 042101 (2010).

\bibitem{ABLS01}
 A. Acin, D. Bruss, M. Lewenstein, and A. Sanpera, 
 Phys. Rev. Lett. {\bf 87}, 040401 (2001).

\bibitem{GHH10}
 A. Gabriel, B. C. Hiesmayr, M. Huber,
 arXiv:1002.2953v1 [quant-ph] (2010).
 
\bibitem{US}
W. Laskowski, M. Markiewicz, T. Paterek and M. \.Zukowski in preparation.

 \end{thebibliography}
\end{document}